\documentclass[a4paper,11pt]{article}

\usepackage{pos}
\usepackage{lipsum} %package to generate placeholder text in the following
\usepackage{caption}
\usepackage{subcaption}

\usepackage{graphicx}
\usepackage{wrapfig}
\usepackage{sidecap}
\usepackage[separate-uncertainty = true]{siunitx}

\newlength{\bibitemsep}
\newlength{\bibparskip}
\let\oldthebibliography\thebibliography
\renewcommand\thebibliography[1]{%
  \oldthebibliography{#1}%
  \setlength{\parskip}{\bibitemsep}%
  \setlength{\itemsep}{\bibparskip}%
}

\title{Deep Learning Based Event Reconstruction for the IceCube-Gen2 Radio Detector}

\ShortTitle{IceCube-Gen2 Radio: Deep Learning Reconstruction}

\author{The IceCube-Gen2 Collaboration \\{\normalsize \normalfont(a complete list of authors can be found at the end of the proceedings)}\\}

\emailAdd{nils.heyer@physics.uu.se}

\abstract{

The planned in-ice radio array of IceCube-Gen2 at the South Pole will provide unprecedented sensitivity to ultra-high-energy (UHE) neutrinos in the EeV range. The ability of the detector to measure the neutrino’s energy and direction is of crucial importance. This contribution presents an end-to-end reconstruction of both of these quantities for both detector components of the hybrid radio array ('shallow' and 'deep') using deep neural networks (DNNs). We are able to predict the neutrino's direction and energy precisely for all event topologies, including the electron neutrino charged-current ($\nu_e$-CC) interactions, which are more complex due to the LPM effect. This highlights the advantages of DNNs for modeling the complex correlations in radio detector data, thereby enabling a measurement of the neutrino energy and direction. We discuss how we can use normalizing flows to predict the PDF for each individual event which allows modeling the complex non-Gaussian uncertainty contours of the reconstructed neutrino direction. Finally, we discuss how this work can be used to further optimize the detector layout to improve its reconstruction performance. 
\vspace{4mm}
{\bfseries Corresponding authors:}
Nils Heyer$^{1*}$, Christian Glaser$^{1}$, Thorsten Glüsenkamp$^{1}$\\
{$^{1}$ \itshape Dept. of Physics and Astronomy, Uppsala University, Box 516, S-75120 Uppsala, Sweden}\\[4mm]

$^*$ Presenter

\ConferenceLogo{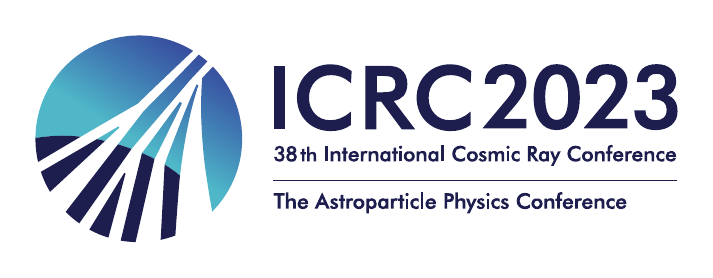}

\FullConference{The 38th International Cosmic Ray Conference (ICRC2023)\\ 26 July -- 3 August, 2023\\ Nagoya, Japan}
}

\begin{document}
\maketitle

\section{The IceCube-Gen2 Radio Array}\label{sec1}

After the IceCube Neutrino Observatory successfully measured the cosmic neutrino flux in the \SI{}{\tera\eV} and low-\SI{}{\peta\eV} range \cite{cosmic}, and identified the first sources of high-energy neutrinos \cite{TXS_0506, NGC_1068}, a larger detector is needed to be sensitive to the rapidly decreasing flux at higher energies. The radio component of IceCube-Gen2 \cite{gen2, hallmann} is planned to instrument \SI{500}{\kilo\meter\squared} of the ice surface with more than 300 radio detector stations and is projected to be sensitive to the neutrino flux at energies into the EeV range. The radio stations can measure radio waves emitted by neutrino interactions via the Askaryan effect \cite{Barwick:2022vqt}. Due to the kilometer-long attenuation length of radio waves in the ice at the South Pole, a single radio station is capable of reaching an effective volume in the order of magnitude of a cubic kilometer in the EeV range. 

\begin{wrapfigure}{l}{0.5\textwidth}
         \centering
         \includegraphics[height=4.5in]{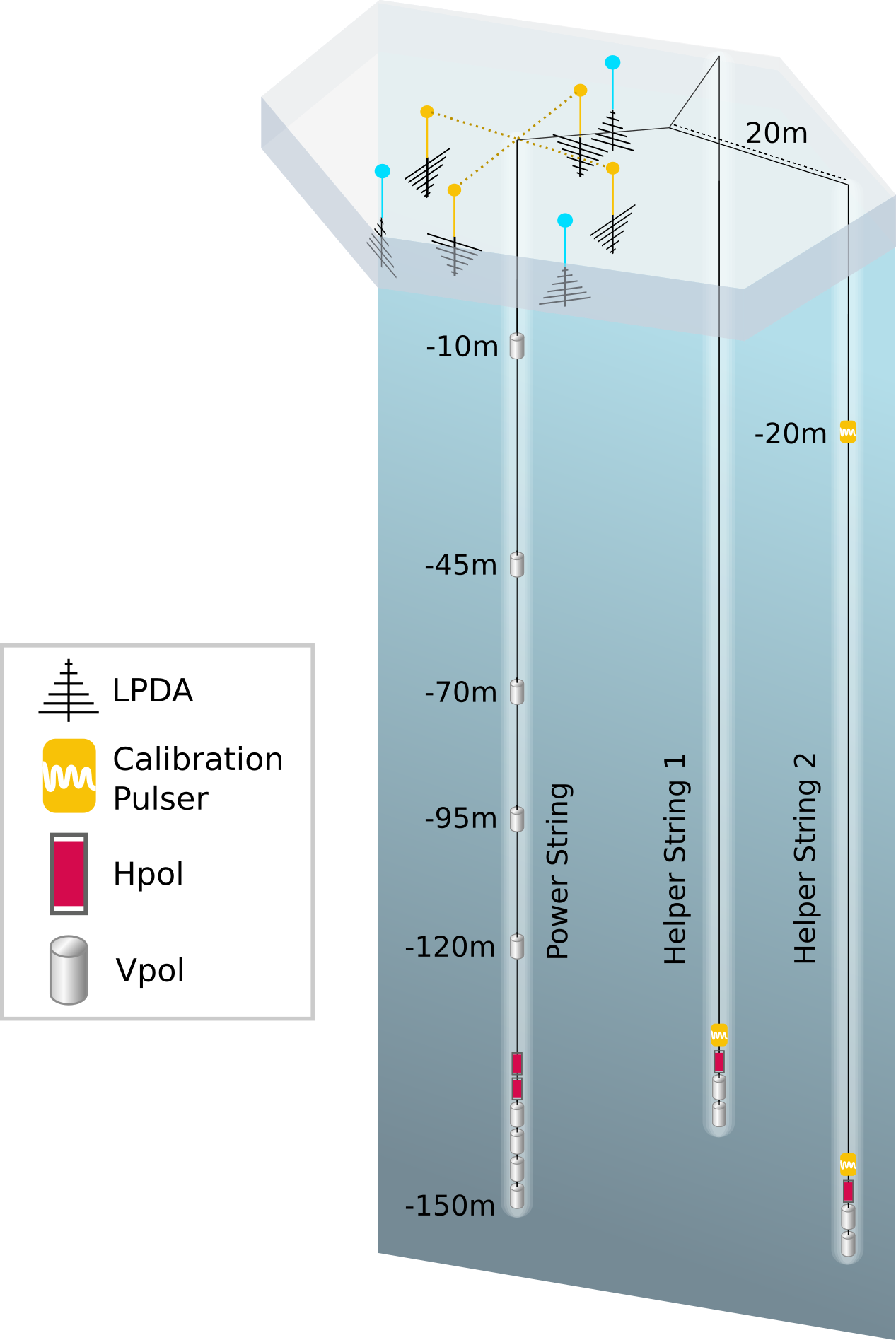}
         \caption{Hybrid station layout \cite{TDR}, combining 8 'shallow' antennas down to \SI{-10}{\meter} and 16 'deep' antennas down to \SI{-150}{\meter}. The triggers for the deep antennas are the four bicone antennas (phased array \cite{phased_array}) at the end of the power string.}
         \label{fig:station}
\end{wrapfigure}

Based on the experiments ARA, ARIANNA, and RNO-G \cite{ARA, ARIANNA1, rnog}, two radio station designs are being developed for the radio component of IceCube-Gen2. The shallow stations are planned to consist of three upward and four downward-facing log-periodic-dipole-array (LPDA) antennas installed just below the snow surface as well as a single \SI{-10}{\meter} deep bicone antenna (Vpol). The hybrid stations include all the components of the shallow stations with twelve additional bicone antennas and four additional slotted cylinder antennas (Hpol) in three \SI{-150}{\meter} deep holes in the ice. The layout of a hybrid station, including the shallow components can be seen in \autoref{fig:station}. The Hpol antennas of the deep components are limited by the size of the drilled holes which results in a more narrow frequency response and less overall gain. Therefore, the direction reconstruction is expected to be more challenging for the deep detector components than for the shallow detector components. For this analysis, we simulate the trigger as proposed for the radio component of IceCube-Gen2 \cite{hallmann}. Indipendent triggers are simulated for the shallow and the deep components of the hybrid stations as a signal triggering the deep antennas often does not trigger the shallow components. For that reason, the analysis of the reconstruction capacity of the station was also split among the shallow and deep components. Furthermore, this analysis focuses only on the reconstruction of signals from a single station and omits the additional information potentially present in neighboring stations which would further improve the reconstruction capabilities. 

\section{Deep Learning Reconstruction}\label{sec2}
A deep-learning-based reconstruction has already been presented previously for shallow stations \cite{Glaser_2023}. Here, we briefly review the results and apply the method to the deep detector component while improving the analysis further. 
The most important component of a deep-learning-based analysis is the dataset used to train and test the models. The dataset used for this analysis was created using NuRadioMC \cite{NuRadioMC}, a Monte Carlo tool capable of simulating the radio signals induced by neutrino interactions. The detector and trigger simulation was performed using NuRadioReco \cite{NuRadioReco}. The dataset for the shallow components (from \cite{Glaser_2023}) was comprised of 40 million events of charged- and neutral-current events but along a non-uniform energy spectrum ranging from \SI[parse-numbers=false]{10^{16}}{\eV} to \SI[parse-numbers=false]{10^{19}}{\eV} with an underrepresentation of low-energy events. The limited and non-uniform energy spectrum had a negative effect on the reconstruction \cite{Glaser_2023} which is why it was improved here for the deep components. The dataset for the deep components is comprised of about 2.1 million events of charged- and neutral-current events along a spectrum uniform in $\log(E)$ ranging from \SI[parse-numbers=false]{10^{16}}{\eV} to \SI[parse-numbers=false]{10^{20.2}}{\eV} of the deposited energy. Even though it is not always possible to identify what interaction type a signal came from, separate analyses were done for neutral- and charged current events to understand these different topologies better.  These simulations were made with our current understanding of the detector and physics processes using the same settings as in \cite{hallmann,Glaser_2023} and the station layout shown in Fig.~\ref{fig:station}.  We acknowledge that systematic uncertainties on the ice model, the antenna response, and the signal chain calibration exist and have to be studied carefully in the future. 

\begin{figure}[tbp]
  \centering
  \includegraphics[height=1.4in]{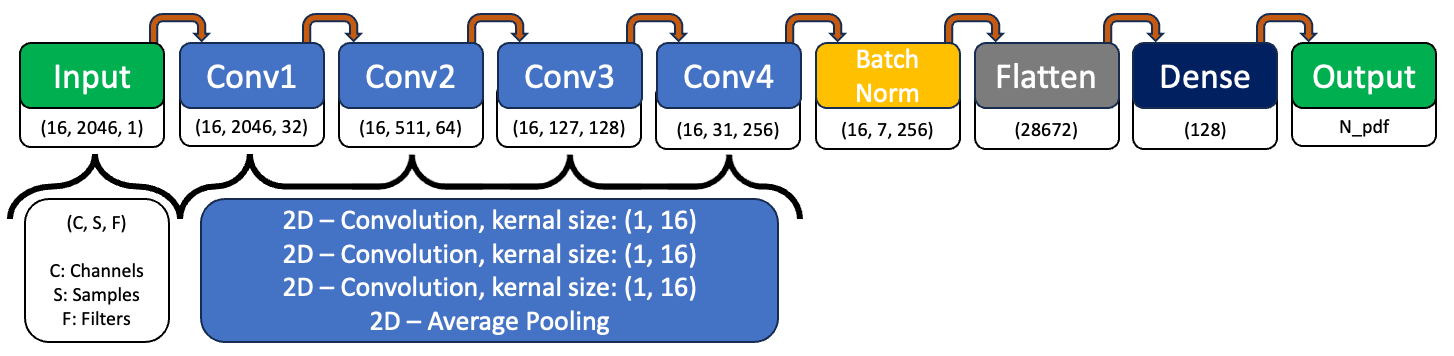}
  \caption{The network architecture used for the analysis of the deep components. The dimensions are indicated below each layer. The output is mapped to a normalizing flow with N\textunderscore pdf parameters.}
  \label{fig:arch}
\end{figure}

The input data to the models is the simulated raw waveforms as measured by the antennas of the shape \emph{antennas} $\times$ \emph{samples} (\emph{5} $\times$ \emph{512} for the shallow components and \emph{16} $\times$ \emph{2046} for the deep components) and the labels are the shower energy induced by the neutrino interaction ($E_{sh}$) and the azimuth and zenith angles of the neutrino direction. Several convolutional layers are then used to analyze the time traces while reducing the length of the traces and increasing the size of the feature dimensions. After batch normalization, the data is flattened and mapped through dense layers (one in the case of the deep components) to the output nodes (for the shallow components \cite{Glaser_2023}) or the output pdf (for the deep components) via a conditional normalizing flow. The architecture of the model used for the deep components can be seen in \autoref{fig:arch}. In the future, we also plan on further improving the architecture of the shallow components with conditional normalizing flows.

A conditional normalizing flow \cite{norm_flow, glüsenkamp} is a way to model an arbitrary conditional PDF. This mapping is a diffeomorphism and it is done via the change-of-variable formula. The parameters of this mapping function can be learned by a neural network. Applying a normalizing flow allows us to predict the PDF for the properties of a neutrino interaction. For the energy, the mean of this one-dimensional PDF can be compared to the true shower energy and its standard deviation gives an estimate of the uncertainty of the prediction. For the direction reconstruction the PDF is mapped onto a two-dimensional sphere where it can be compared to the true direction and the area of the uncertainty contours can be calculated. The normalizing flows used for this analysis were implemented using the \emph{jammy\textunderscore flows} library \cite{jammy}. The energy reconstruction used two gaussianization-flows and a multivariate-normal-flow while the direction reconstruction used a exponential-map-2d-sphere-flow.

\section{Results}\label{sec3}

In the following, we will shortly present the results obtained from a deep learning analysis for the resolution of the shallow components before going into detail about the analysis for the deep station components. The conclusion will offer a comparison between the two analyses.

\subsection{Shallow Station Reconstruction}

Using a deep neural network, simulated pulses were analyzed for the shallow station components \cite{Glaser_2023}. The resolution of the energy was determined with a standard deviation of $\sigma \approx 0.3$ in $\log_{10}(E_{sh})$. For the first time, predictions of the neutrino direction for all event topologies including the complicated electron neutrino charged-current ($\nu_e$-CC) interactions were made possible for the shallow station components. 
The obtained angular resolution shows a narrow peak at $\mathcal{O}$(\SI{1}{\degree}) with extended tails that push the 68\% quantile for non-$\nu_e$-CC (resp. $\nu_e$-CC interactions) to $\SI{4}{\degree} (\SI{5}{\degree})$. Due to the non-uniform energy spectrum used to train the network, the resolution decreased significantly at low and high energies.

\subsection{Deep Station Reconstruction}

After the analysis of the shallow stations, it became clear that the energy spectrum had to be extended and have a uniform shape to avoid a bias at low and high energies. Additionally, conditional normalizing flows were introduced allowing for event-by-event uncertainty predictions.

%\subsubsection{Energy Reconstruction}

\begin{figure}[tbp]
  \centering
  \includegraphics[height=2.5in]{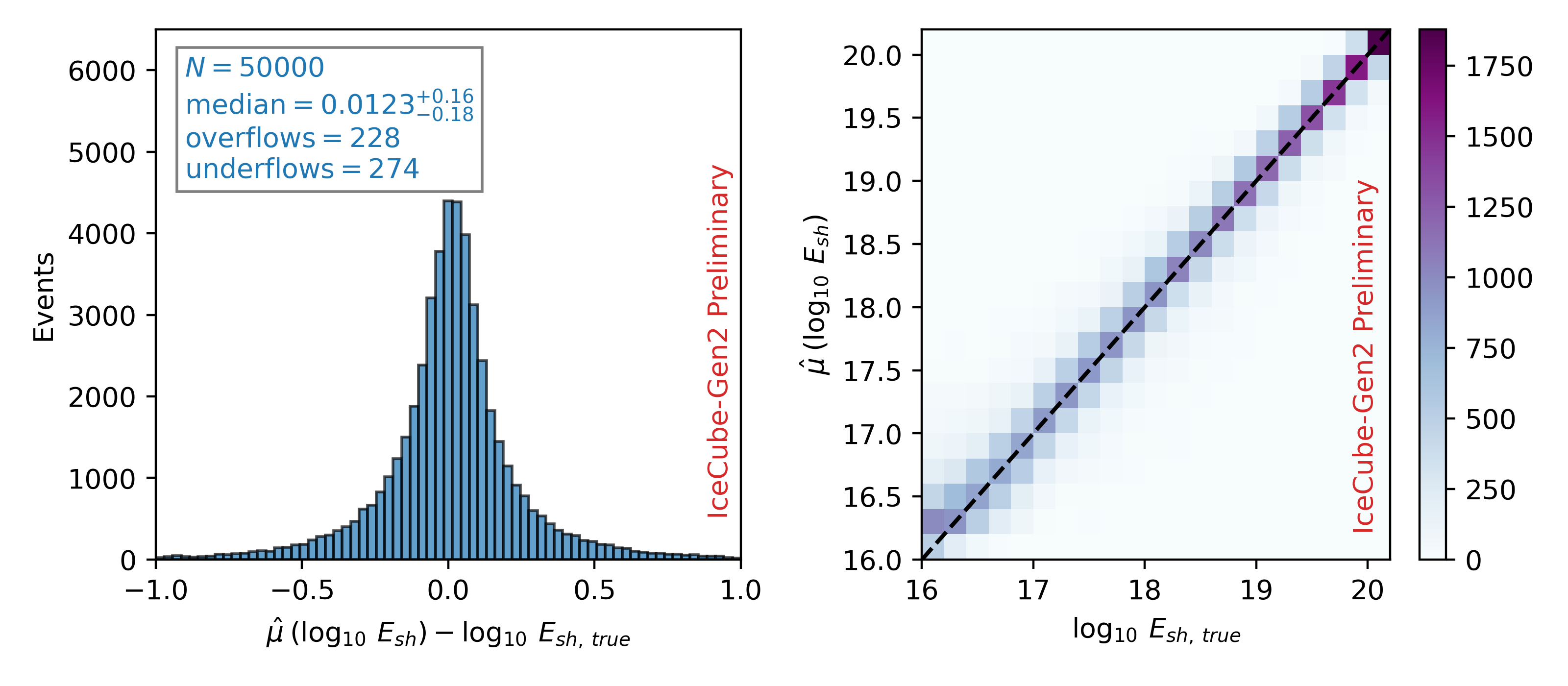}
  \caption{A comparison of the true shower energy and the predicted shower energy by the network over all events in the test data set. The left plot shows the difference between the mean of the predicted distribution and the true shower energy. The $16\%$ and $84\%$ percentiles shown behind the median indicate the resolution of the model on the full dataset.}
  \label{fig:ggt}
\end{figure}

As the emitted radio pulses are only affected by the energy the neutrino deposits in the ice rather than the neutrino energy itself, the networks were trained on the shower energy, which includes all energy deposited after a neutrino interaction. This includes hadronic and electromagnetic showers. Therefore, the results for the neutral-current events have to be folded with the inelasticity of the neutrino interaction to calculate the neutrino energy while the charged-current interactions of electron-neutrinos deposit all the neutrino energy as shower energy. 

\autoref{fig:ggt} shows the results of the energy reconstruction for the full energy range. The left plot displays the distribution when subtracting the true shower energy from the mean of the predicted PDF of the shower energy for every event. The median is centered around zero (0.0123) and the values of the $16\%$ percentile (-0.18) and the $84\%$ percentile (+0.16) are the energy resolution of the analysis. The right plot displays a comparison of the true shower energy and the predicted shower energy as a function of shower energy. It is visible that the reconstruction gets better the higher the shower energy gets as the 2d-histogram is centered more around the diagonal. At energies below $E_{sh} \approx \SI[parse-numbers=false]{10^{16.5}}{\eV}$ the network is overpredicting the shower energy similarly to the analysis for the shallow stations, but with a much smaller effect.

\begin{figure}[tbp]
  \centering
  \includegraphics[height=2.5in]{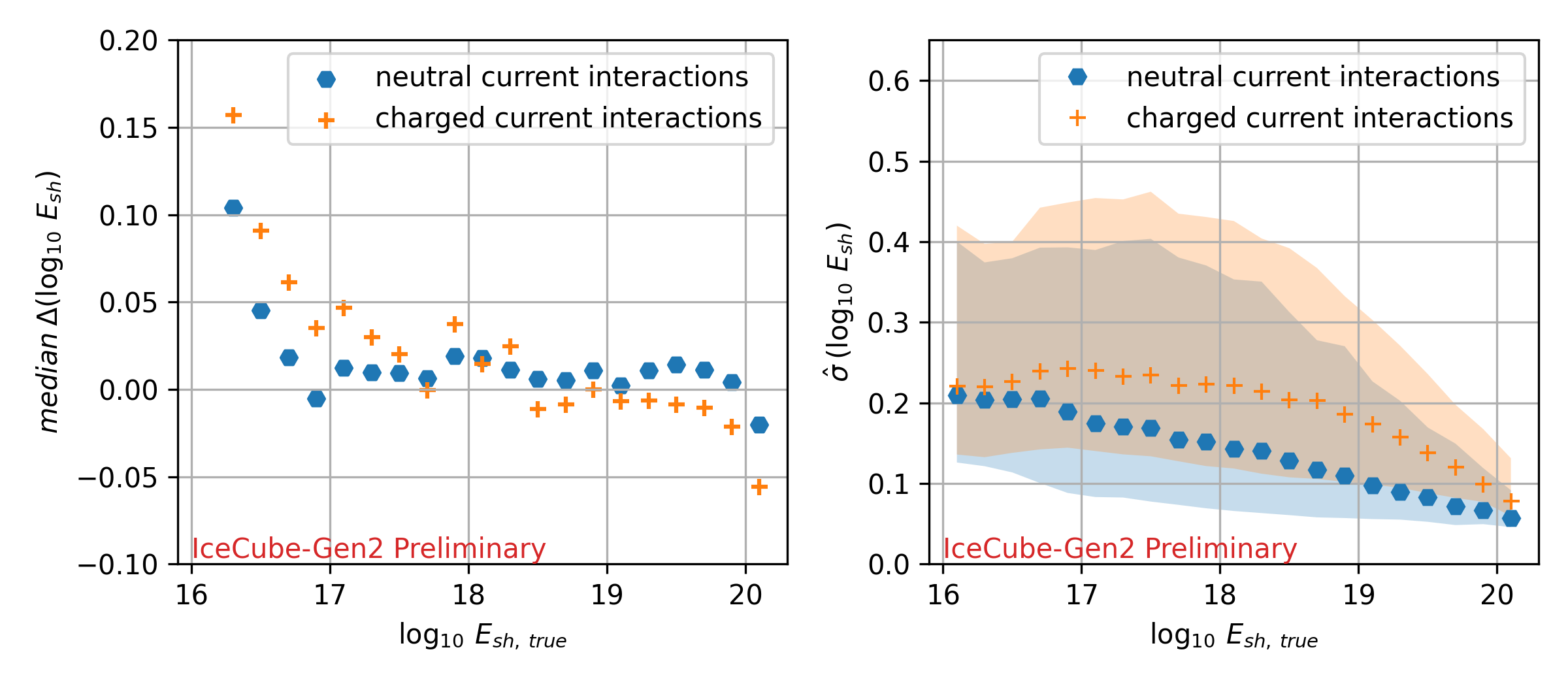}
  \caption{The energy bias (left) and resolution (right) as a function of shower energy. The median on the left plot corresponds to the left plot in \autoref{fig:ggt} for different energy ranges. For the right plot, the standard deviation of the PDF for every predicted event was collected for every energy bin. The median of the resulting distributions is indicated with markers while the $16\%$ percentile and the $84\%$ percentile are indicated with the shaded regions. }
  \label{fig:nccc}
\end{figure}

\autoref{fig:nccc} shows the resolution of the energy reconstruction per energy bin. It is visible that at energies below $E_{sh} \approx \SI[parse-numbers=false]{10^{17}}{\eV}$ the median has a bias. However, in the most relevant range between $E_{sh} \approx \SI[parse-numbers=false]{10^{17}}{\eV} - \SI[parse-numbers=false]{10^{19}}{\eV}$ the median is centered around zero. The bias towards low energies can likely be reduced by extending the dataset towards lower energies. The predicted uncertainties show a clear energy dependence where the values drop from about 0.2 at the lowest shower energies to about 0.05 at the highest shower energies. However, it has to be stated that the coverage (a measure of how the true shower energy values compare to the uncertainty contours) for these uncertainties, showed an up to $10\%$ deviation from the expected coverage, which indicates an underestimation of the uncertainty prediction. This deviation in the coverage indicates that the network can still be further optimized to yield more accurate uncertainty predictions for the shower energy reconstruction. 

%\subsubsection{Direction Reconstruction}

\begin{SCfigure}
  \centering
  \caption{Mollweide projection of the full sky in local coordinates. The arrival direction is affected by the opaqueness of the Earth which results in most events coming from above the horizon. Displayed are two examples of the event-by-event direction reconstruction. Top: A well-reconstructed event at \SI[parse-numbers=false]{10^{17.77}}{\eV} with an almost Gaussian uncertainty contour. Bottom: A less well-reconstructed event at \SI[parse-numbers=false]{10^{17.13}}{\eV} with a larger non-gaussian uncertainty contour. The $68\%$ uncertainty contour is indicated with a dashed line and the $95\%$ uncertainty contour is indicated with a dotted line. The true direction that the model is trying to reconstruct is indicated with a red cross.} 
  \label{fig:examples}
    %\begin{minipage}[c]{0.7\textwidth}
    \begin{minipage}[c]{0.68\textwidth}
    \includegraphics[width=\linewidth]{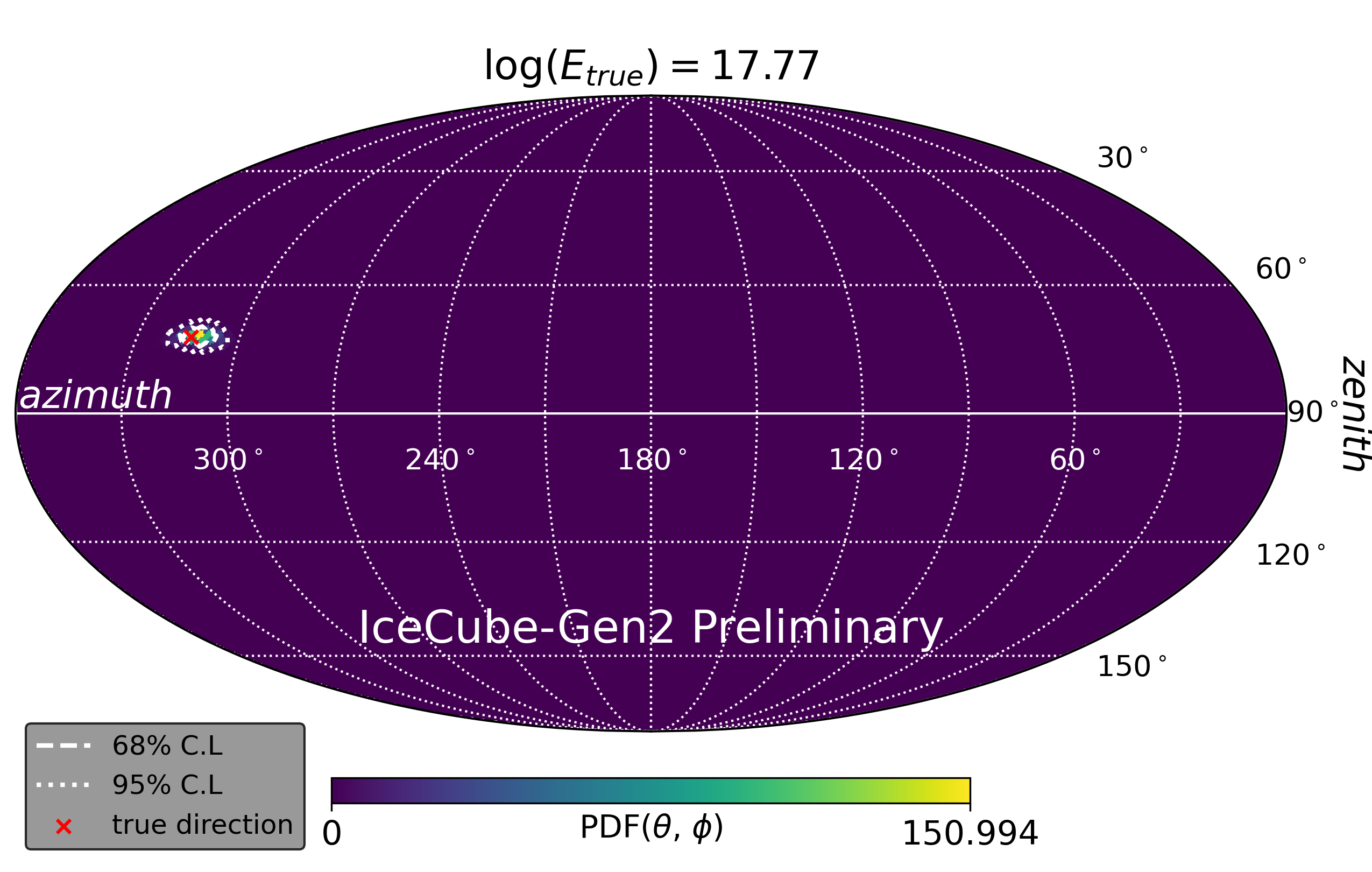}
    \includegraphics[width=\linewidth]{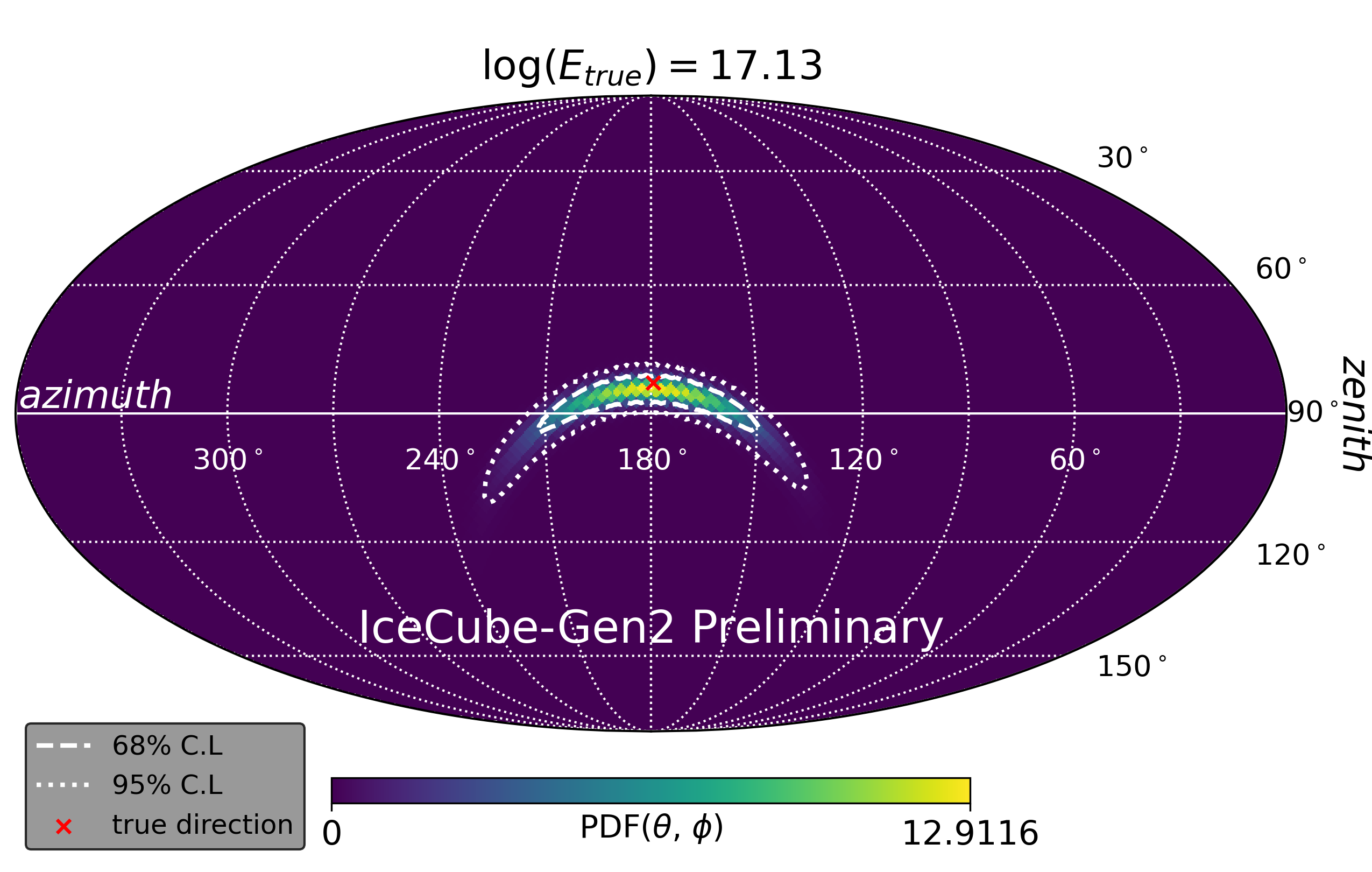}
\end{minipage}
\end{SCfigure}

Reconstructing the direction using normalizing flows involves different methods when comparing the network prediction with the true label than the methods previous analyses used. Previous reconstructions predicted a single direction where the angle between the predicted and the true direction was used to estimate the uncertainty. As we are now predicting a non-Gaussian-shaped PDF rather than a single vector other methods are needed. First, the coverage of the PDF has to be checked to see if the true directions correspond to the predicted PDFs. Second, the spread of the PDFs has to be estimated which can be done by calculating the entropy of the PDF on the sphere or by calculating the area of the uncertainty contours for every single event. A space-angle difference is only statistically meaningful with the mean of the predicted PDF if the predicted uncertainty contours are Gaussian-shaped. So far, we included only hadronic showers in the analysis of the deep station component, while the analysis for electromagnetic showers is ongoing.

Before deep learning was used, the direction reconstruction was done using a forward folding technique \cite{NuRadioReco,ARIANNADirectionICRC2021,Ilse, Sjoerd}. The deep-learning technique, including normalizing flows as presented in this contribution, allows for event-by-event predictions of the PDF and therefore also uncertainty contours. Two examples of these predicted PDFs compared with the true direction of the event can be seen in \autoref{fig:examples}.

\begin{figure}
     %\centering
     \begin{subfigure}[b]{0.49\textwidth}
  \includegraphics[height=2.65in]{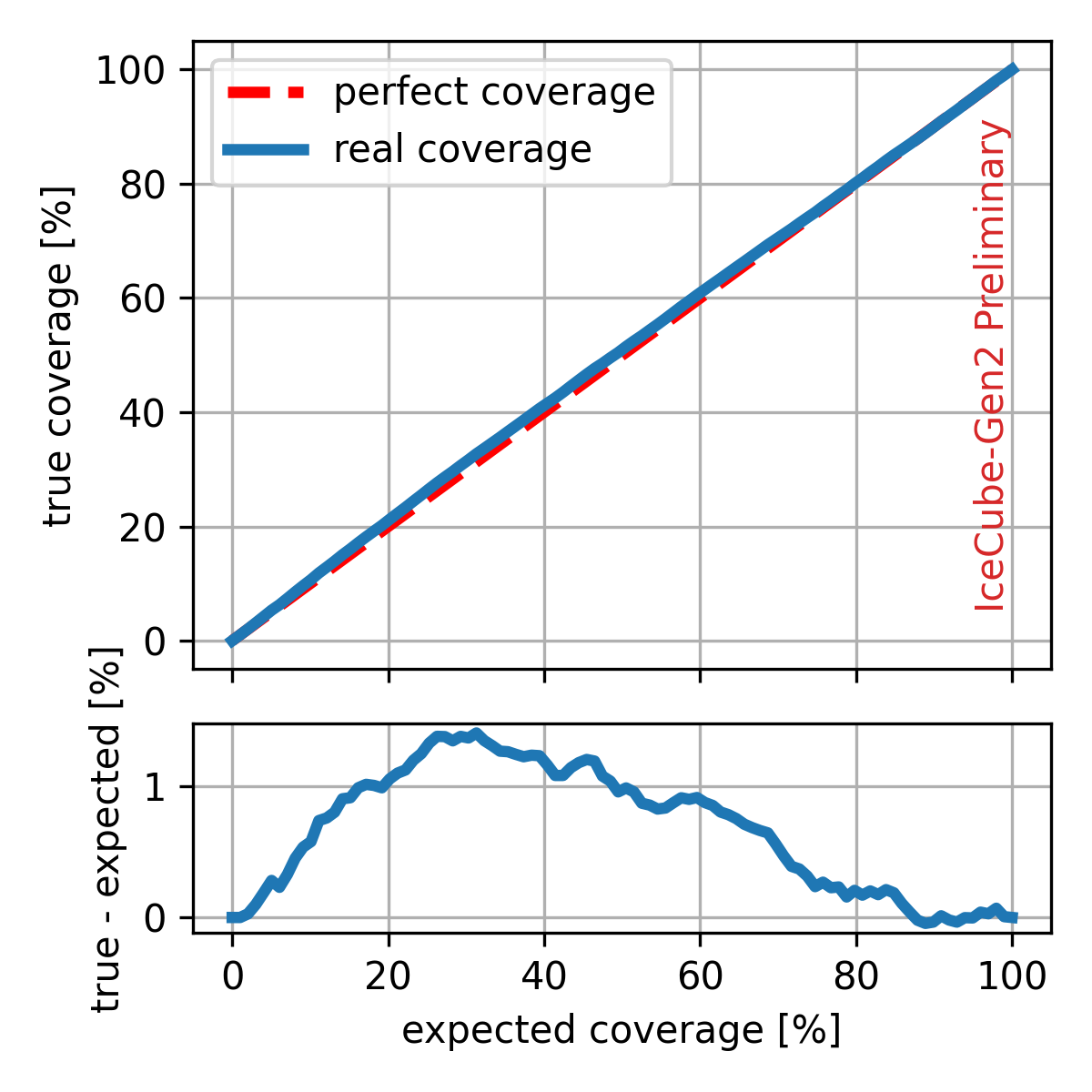}
     \end{subfigure}
\hspace*{-1.0cm}
     \begin{subfigure}[b]{0.49\textwidth}
  \includegraphics[height=2.65in]{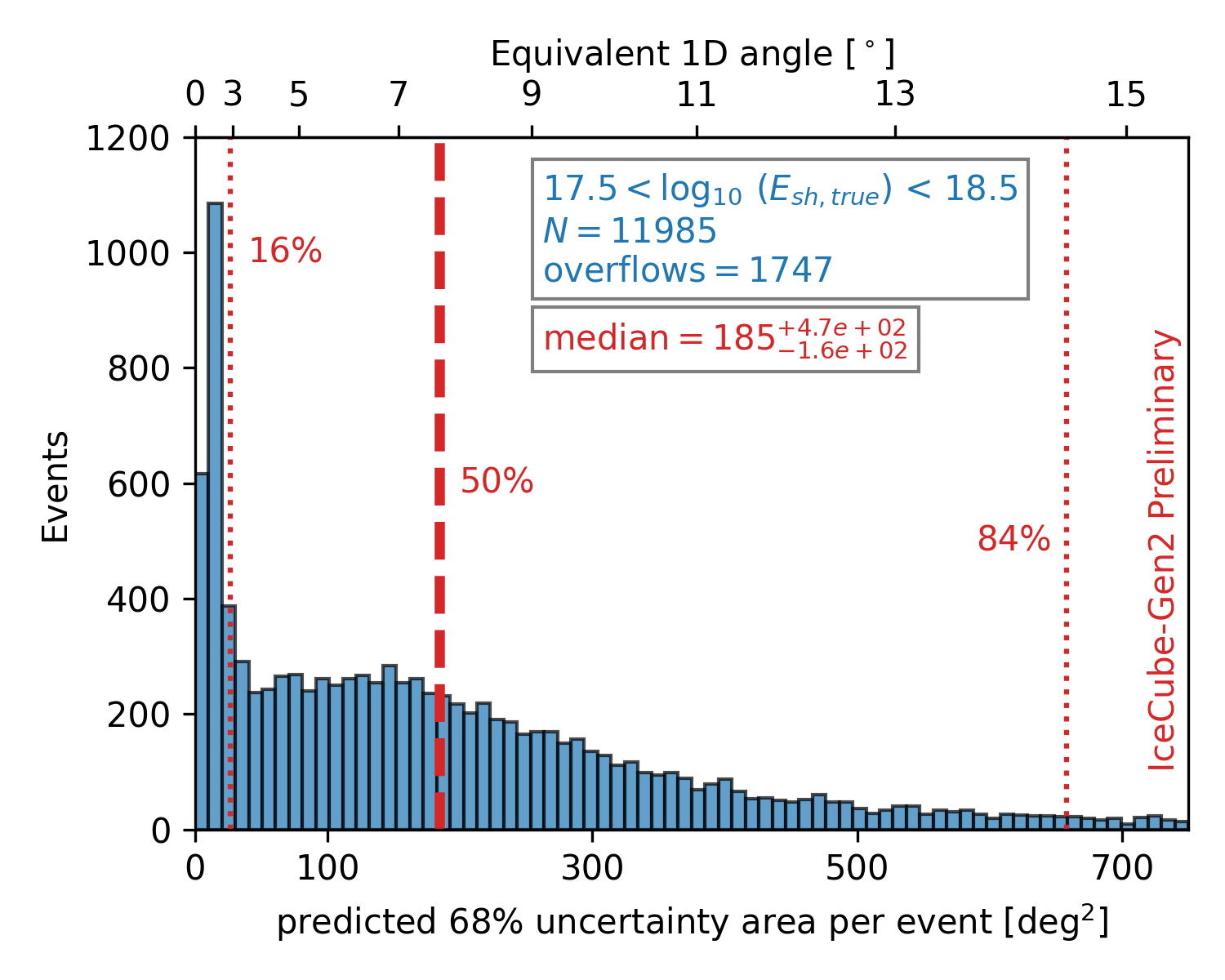}
     \end{subfigure}
        \caption{Left: The coverage of all test events for the direction reconstruction. The top plot shows how the real coverage compares to a perfect estimation of the coverage. The bottom plot shows how much the true coverage deviates from the expected coverage. Right: The areas of the $68\%$ uncertainty contours of every event in the shower energy range from $E_{sh} \approx \SI[parse-numbers=false]{10^{17.5}}{\eV} - \SI[parse-numbers=false]{10^{18.5}}{\eV}$. The space-angle difference equivalent for a Gaussian distribution with the same area is shown as the top x-axis. The $16\%$, $50\%$, and $84\%$ percentiles are indicated with red lines to evaluate the shape of the area distribution.}
        \label{fig:cov_area}
\end{figure}

The coverage \cite{glüsenkamp} of the direction reconstruction, displayed in \autoref{fig:cov_area} (left), shows a good agreement between the true and the expected coverage with the maximum deviations around $1\%$. This indicates that the predicted uncertainty contours can be trusted to a high degree when applying this network to a new, unknown event of the same class as the training class. \autoref{fig:cov_area} (right) shows the distribution of the areas of the predicted two-dimensional $68\%$ uncertainty contours with the equivalent space-angle difference in the most relevant energy range ($E_{sh} \approx \SI[parse-numbers=false]{10^{17.5}}{\eV} - \SI[parse-numbers=false]{10^{18.5}}{\eV}$). It indicates that there are about $16\%$ of high-quality events which can be reconstructed with an equivalent space-angle difference of less than $3^\circ$. These high-quality samples can be selected directly from the size of the predicted uncertainty contours. After that, the distribution flattens and the median sits at about $7.5^\circ$. The distribution also has a very long tail with about $15\%$ of the events not even fitting on the plot, indicating very low-quality events.

\section{Conclusion}\label{sec4}

For this contribution, a deep-learning-based reconstruction was explored for energy and direction reconstruction of neutrinos detected via in-ice emitted radio signals. The analysis of the shallow station components resulted in an energy resolution of $\sigma \approx 0.3$ in $\log_{10}(E_{sh})$) while the analysis on the deep components resulted in an energy resolution of $\sigma \approx 0.2$ in $\log_{10}(E_{sh})$). This improvement can likely be attributed to the uniform shape of the training data set used for the deep components. Previous findings using traditional approaches were also able to reconstruct the neutrino energy within the natural inelasticity limit of a factor of two (or 0.3 in $\log_{10}(E)$) but were reliant on analysis cuts to filter out low-quality events \cite{ARIANNA_energy, RNOG_energy}. The direction reconstruction of neutral-current events showed an average resolution of $4^\circ$ for the shallow components and $7.5 ^\circ$ for the deep components. This discrepancy very likely comes from the less sensitive horizontally polarized antennas used by the deep station components and is compatible with previous findings using traditional reconstruction techniques \cite{ARIANNA_forward_folding,ARIANNADirectionICRC2021,Ilse}. The improved methods laid out for the deep components in this contribution will also be applied to the shallow components and further improvements can be expected. Several aspects of the analysis of the deep components also indicate that further improvements in the network model could result in an even better performance. These improvements could include changes to the convolutional or dense layer architecture as well as training a model using neutral- and charged-current interactions in the same data set. Perhaps training a model which can predict shower energy and neutrino direction at the same time would also increase the performance of the neural network.

% Bibtex references:
\bibliographystyle{ICRC}
\bibliography{references}

\clearpage

%The following list of authors, affiliations and funding agencies will be updated at the day of submission. The following template is a placeholder generated via https://authorlist.icecube.wisc.edu/icecube on March 25, 2023 and will be updated.
\section*{Full Author List: IceCube-Gen2 Collaboration}

\scriptsize
\noindent
R. Abbasi$^{17}$,
M. Ackermann$^{76}$,
J. Adams$^{22}$,
S. K. Agarwalla$^{47,\: 77}$,
J. A. Aguilar$^{12}$,
M. Ahlers$^{26}$,
J.M. Alameddine$^{27}$,
N. M. Amin$^{53}$,
K. Andeen$^{50}$,
G. Anton$^{30}$
% ...
\\
\\
% ...
$^{12}$ Universit{\'e} Libre de Bruxelles, Science Faculty CP230, B-1050 Brussels, Belgium \\
% ...
$^{17}$ Department of Physics, Loyola University Chicago, Chicago, IL 60660, USA \\
% ...
$^{22}$ Dept. of Physics and Astronomy, University of Canterbury, Private Bag 4800, Christchurch, New Zealand \\
% ...
$^{26}$ Niels Bohr Institute, University of Copenhagen, DK-2100 Copenhagen, Denmark \\
$^{27}$ Dept. of Physics, TU Dortmund University, D-44221 Dortmund, Germany \\
% ...
$^{30}$ Erlangen Centre for Astroparticle Physics, Friedrich-Alexander-Universit{\"a}t Erlangen-N{\"u}rnberg, D-91058 Erlangen, Germany \\
% ...
$^{47}$ Dept. of Physics and Wisconsin IceCube Particle Astrophysics Center, University of Wisconsin{\textendash}Madison, Madison, WI 53706, USA \\
% ...
$^{50}$ Department of Physics, Marquette University, Milwaukee, WI, 53201, USA \\
% ...
$^{53}$ Bartol Research Institute and Dept. of Physics and Astronomy, University of Delaware, Newark, DE 19716, USA \\
% ...
$^{76}$ Deutsches Elektronen-Synchrotron DESY, Platanenallee 6, 15738 Zeuthen, Germany  \\
$^{77}$ Institute of Physics, Sachivalaya Marg, Sainik School Post, Bhubaneswar 751005, India \\
% ...

\subsection*{Acknowledgements}

\noindent
The authors gratefully acknowledge the support from the following agencies and institutions:
USA {\textendash} U.S. National Science Foundation-Office of Polar Programs,
U.S. National Science Foundation-Physics Division,
U.S. National Science Foundation-EPSCoR,
Wisconsin Alumni Research Foundation,
Center for High Throughput Computing (CHTC) at the University of Wisconsin{\textendash}Madison,
Open Science Grid (OSG),
Advanced Cyberinfrastructure Coordination Ecosystem: Services {\&} Support (ACCESS),
Frontera computing project at the Texas Advanced Computing Center,
U.S. Department of Energy-National Energy Research Scientific Computing Center,
Particle astrophysics research computing center at the University of Maryland,
Institute for Cyber-Enabled Research at Michigan State University,
and Astroparticle physics computational facility at Marquette University;
Belgium {\textendash} Funds for Scientific Research (FRS-FNRS and FWO),
FWO Odysseus and Big Science programmes,
and Belgian Federal Science Policy Office (Belspo);
Germany {\textendash} Bundesministerium f{\"u}r Bildung und Forschung (BMBF),
Deutsche Forschungsgemeinschaft (DFG),
Helmholtz Alliance for Astroparticle Physics (HAP),
Initiative and Networking Fund of the Helmholtz Association,
Deutsches Elektronen Synchrotron (DESY),
and High Performance Computing cluster of the RWTH Aachen;
Sweden {\textendash} Swedish Research Council,
Swedish Polar Research Secretariat,
Swedish National Infrastructure for Computing (SNIC),
and Knut and Alice Wallenberg Foundation;
European Union {\textendash} EGI Advanced Computing for research;
Australia {\textendash} Australian Research Council;
Canada {\textendash} Natural Sciences and Engineering Research Council of Canada,
Calcul Qu{\'e}bec, Compute Ontario, Canada Foundation for Innovation, WestGrid, and Compute Canada;
Denmark {\textendash} Villum Fonden, Carlsberg Foundation, and European Commission;
New Zealand {\textendash} Marsden Fund;
Japan {\textendash} Japan Society for Promotion of Science (JSPS)
and Institute for Global Prominent Research (IGPR) of Chiba University;
Korea {\textendash} National Research Foundation of Korea (NRF);
Switzerland {\textendash} Swiss National Science Foundation (SNSF);
United Kingdom {\textendash} Department of Physics, University of Oxford.

\end{document}